\documentstyle[aps,epsfig]{revtex}

\begin{document}

\title{Evidence of large nuclear deformation of $^{32}$S$^{*}$ formed 
in $^{20}$Ne + $^{12}$C reaction}

\author{Aparajita Dey, S.~Bhattacharya, C.~Bhattacharya, 
K.~Banerjee, T.~K.~Rana, \\
S.~Kundu, S.~Mukhopadhyay, D.~Gupta, R.~Saha}

\address{ Variable Energy Cyclotron Centre, 1/AF Bidhan Nagar, 
Kolkata - 700064, India}

\maketitle

\begin{abstract}
Deformations of hot composite $^{32}$S$^{*}$ formed in the reaction
$^{20}$Ne ($\sim$ 7 -- 10 MeV/nucleon) + $^{12}$C have been
estimated from the respective inclusive $\alpha$-particle
evaporation spectra. The estimated deformations for $^{32}$S$^{*}$
have been found to be much larger than the `normal' deformations of
hot, rotating composites at similar excitations. This further
confirms the formation of highly deformed long-lived configuration
of $^{20}$Ne + $^{12}$C at high excitations ($\sim$ 70 -- 100 MeV)
--- which was recently indicated from the analysis of the complex 
fragment emission data for
the same system. Exclusive $\alpha$-particle evaporation spectra
from the decay of hot composite $^{32}$S$^{*}$ also show similar
behaviour.
\end{abstract}


\section{Introduction}
The phenomena of clustering and large deformations in light N = Z
nuclei has recently evoked a lot of interest and a few attempts have
been made to study the characteristics of such light, highly
deformed systems \cite{hori03,hori04,von,thum}. Formation of 
long-lived, highly
deformed dinuclear configuration has been observed in low energy,
light heavy-ion induced reactions involving $\alpha$-like nuclei
\cite{beck}. These highly deformed systems are generally believed to
be formed through nuclear orbiting mechanism. Orbiting can be
described in terms of the formation of a long-lived dinuclear
complex which acts as a ``doorway'' state to fusion with a strong
memory of the entrance channel - though the dynamics is not yet
completely understood. It is particularly of interest to know how
the deformed dinuclear shape (and vis-\'{a}-vis the orbiting
process) evolves with the increase in excitation energy. Various
theoretical calculations also predict the formation of such highly
deformed shapes of light mass (A $<$ 100) N = Z nuclei at higher
excitation energy and angular momentum \cite{aberg,zhang}. Recently
it has been reported that for $^{20}$Ne ($\sim$ 7 -- 10 MeV/nucleon)
+ $^{12}$C reactions, there is large enhancement in the yield of the
Carbon and Boron fragments \cite{ref9}, which may be indicative of
the survival of orbiting at these high excitation energies ($\sim$
70 -- 100 MeV); As orbiting is associated with the formation of
highly deformed dinuclear configuration, it will be interesting to
investigate into the quantitative deformation of the hot composites
produced in the same reactions independently through other available
experimental tools.

Light charged particles (LCP; Z $\leq$ 2) spectroscopy is now-a-days
routinely used to study the statistical properties of the hot
rotating nuclei \cite{ref1}. In the case of the compound nuclei at
moderate energies and angular momenta, such as those produced with
light ion projectiles, the experimental spectra are well reproduced
in terms of statistical models which employ the standard optical
model transmission coefficients. However, in the case of heavy ion
induced fusion reactions, {\it i.e.}, at high excitation energies
and angular momenta, the energy spectra of evaporated light charged
particles are no longer consistent with the standard statistical
model predictions \cite{ref1,ref3,ref4,viesti,ref6,ref7,rous}. 
The discrepancy
is believed to be due to the deformation of the composite system at
high excitation energy and angular momentum and proper modification
of the emission barrier and/or the level density is needed to
incorporate the effect of deformation 
\cite{ref4,viesti,ref6,ref7,rous,ref8}.
Apart from that, entrance channel dynamics may also affect the
evaporative decay if the composite forms a long lived, highly
deformed configuration (as observed in the reactions involving light
$\alpha$-like nuclei) with lifetime comparable to the evaporation
time scale. Thus the LCP's in general (and $\alpha$'s in particular,
as they are emitted predominantly at the initial stages of the decay
cascade) can provide important clue about the nuclear deformation at
high excitation energy and spin. This prompted us to explore the
quantitative deformation of $^{20}$Ne + $^{12}$C dinuclear system in
the incident energy range $\sim$ 7 -- 10 MeV/nucleon from the
respective $\alpha$-evaporation spectra.

In the present work, we have populated the hot composite nucleus
$^{32}$S$^{*}$ in the excitation energy range $\sim$ 73.0 MeV to
94.0 MeV through $^{20}$Ne + $^{12}$C reaction at different
bombarding energies. Inclusive energy and angular distributions of
all the light charged particles have been measured. In-plane
coincidence of light charged particles with both evaporation
residues (ER) and intermediate mass fragments (IMF) has been
measured at one incident energy (158 MeV) for comparison of
inclusive and exclusive LCP emission spectra. Inclusive energy and
angular distributions of the LCPs emitted from the hot composite
formed in $^{20}$Ne (158 MeV) + $^{27}$Al reactions have also been
measured. The deformations of the hot composites formed in the above
reactions have been estimated from the study of the respective
$\alpha$-evaporation spectra. It is found that the deformations of
hot composites formed in $^{20}$Ne + $^{12}$C reaction are much
larger than the `normal' deformations, which hot, rotating compound
nuclei ({\it i.e.}, $^{31}$P$^{*}$, $^{47}$V$^{*}$, formed in
$^{19}$F + $^{12}$C \cite{ref1}, $^{20}$Ne + $^{27}$Al reactions,
respectively) may undergo at similar excitations. This further
corroborates the conjecture of the formation of highly deformed,
long-lived configuration of $^{20}$Ne + $^{12}$C system at high
excitations, which was earlier indicated from the enhancement of
fragment yield from the same system at these energies.

The paper has been arranged as follows. The
experimental procedures are described in the Sec.~II. The analysis 
of the data and the results are discussed in Sec.~III. Finally, the 
summary and concluding remarks are given in Sec.~IV.

\section{Experimental details}
The experiments were performed with accelerated $^{20}$Ne ion beams
of energies 145, 158, 170, 180 and 200 MeV, respectively, from the
Variable Energy Cyclotron at Kolkata. The targets used were $\sim$
550 $\mu$g/cm$^{2}$ self-supporting $^{12}$C and $\sim$ 500
$\mu$g/cm$^{2}$ $^{27}$Al foil. The intermediate mass fragments and
evaporation residues have been detected using two solid state
$\Delta$E - E [$\sim$ 10 $\mu$m Si(SB) $\Delta$E, $\sim$ 300 $\mu$m
Si(SB) E] telescopes (THI) mounted on one arm of the 91.5 cm
scattering chamber. The light charged particles have been detected
using two solid state thick detector telescopes [$\sim$ 40, 100
$\mu$m Si(SB) $\Delta$E, $\sim$ 5 mm Si(Li) E] (TLI) mounted on the
other arm of the scattering chamber. Typical solid angles subtended
by the telescopes were 0.33 msr, 0.19 msr, 0.74 msr and 0.53 msr,
respectively. The telescopes were calibrated using elastically
scattered $^{20}$Ne ion from Au, Al and C targets and $^{228}$Th
$\alpha$-source. Absolute energy calibrations of the E and $\Delta$E
detectors for each telescopes were done separately using standard
kinematics and energy-loss calculations. The measured energies have
been corrected for the energy losses at the target by incorporating
a single average thickness correction for each fragment energy. The
low-energy cutoffs thus obtained were typically $\sim$ 3.0 MeV for
proton and $\sim$ 10.2 MeV for $\alpha$-particles in TLI telescopes.
In THI telescopes, the cutoffs were $\sim$ 12.3 MeV for Boron and
$\sim$ 19.5 MeV for Oxygen. Well separated bands corresponding to
elements having atomic numbers up to Z = 13 have been identified.

Inclusive energy distributions for light charged particles (Z = 1,2)
have been measured in the angular range 10$^{o}$ to 50$^{o}$, 
in steps of 2.5$^{o}$, for
$^{20}$Ne + $^{12}$C at all the bombarding energies and for $^{20}$Ne +
$^{27}$Al at 158 MeV. We have also measured the in-plane coincidence
of light charged particles with intermediate mass fragments and
evaporation residues for $^{20}$Ne + $^{12}$C at 158 MeV for the
comparison of inclusive and exclusive spectra.

\section{Results and Discussions}
\subsection{Energy distribution}
\begin{figure} [h]

{\epsfig{file=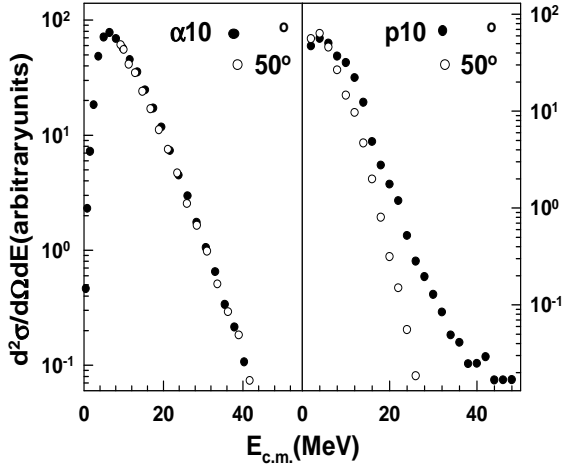,width=8cm,height=7cm.}}

\caption{Comparison of energy spectra of $\alpha$-particles (left), and, proton (right) 
at two angles for the reaction $^{20}$Ne (E$_{lab}$ = 200 MeV) + $^{12}$C  .}
\label{alp1}
\end{figure}

Typical inclusive energy spectra of $\alpha$-particles have been shown 
in Fig.~\ref{alp1}~(left) for the bombarding energy 200 MeV. The energy spectrum 
at $\theta_{lab}$ = 10$^{o}$ (filled circles) is compared with that at 
50$^{o}$ (open circles). It is clearly seen from the figure that the 
slopes of the two spectra are matching well. 

\begin{figure} [h]

{\epsfig{file=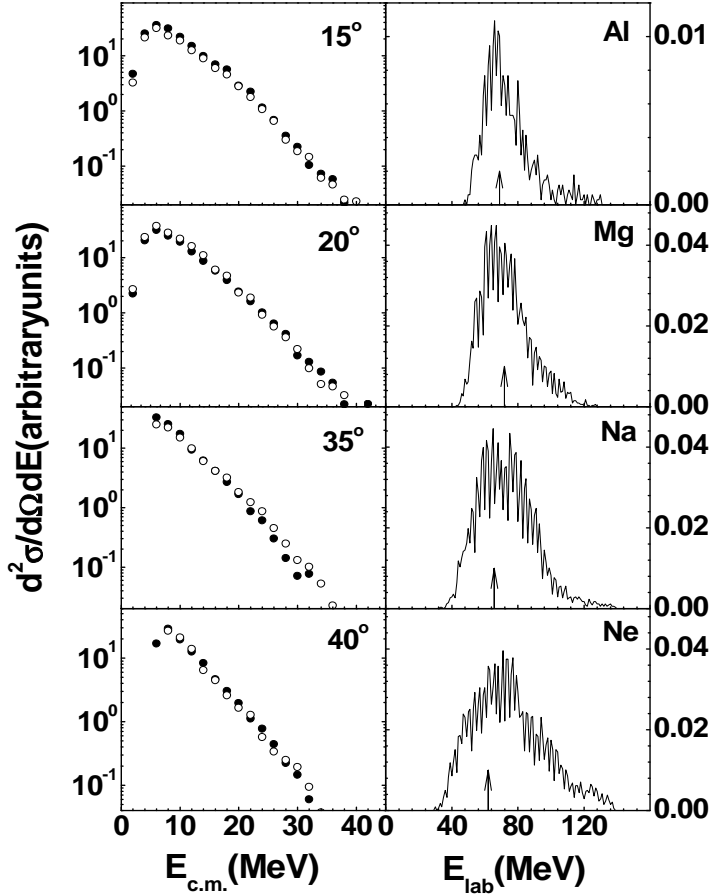,width=10cm,height=12.5cm.}}

\caption{ Left: Comparison between inclusive (filled circles) and
exclusive (open circles, measured in coincidence with all ER's emitted at 
$\theta_{lab}$ = 10$^{o}$) spectra for $\alpha$-particles at E$_{lab}$ = 
158 MeV at different angles. Cross-sections were
normalized at one point for comparison. Right: Evaporation residues spectra 
at $\theta_{lab}$ = 10$^{o}$ measured in coincidence with LCP's ({\it see text}). 
The arrow indicates the energy corresponding to the residue velocity 
$\simeq$ $v_{CN} \cos \theta_{lab}$ for the respective ER.}
\label{alp2}
\end{figure}

The shapes of the inclusive $\alpha$-particle spectra have been
compared with the respective exclusive spectra (coincident
with ER, Z = 10 -- 13) at $\theta_{lab}$ = 15$^{o}$, 20$^{o}$, 35$^{o}$ 
and 40$^{o}$ in Fig.~\ref{alp2}~(left pannel). It is found that the 
shapes of the inclusive and
exclusive $\alpha$-particle spectra are matching well for all the angles. 

In heavy-ion collisions with bombarding energies well above the Coulomb 
barrier, a significant fraction of the LCPs (particularly protons) are 
known to be emitted at forward angles in the early stage of the reaction 
\cite{awes}. The forward angle LCP energy distributions have strong 
enhancement in the high-energy part when compared with those of  
evaporation spectra expected in compound nuclear reactions. This extra 
contribution in forward direction is generally a signature of 
pre-compound emission, and it becomes insignificant at 
$\theta_{lab}$ $>$ 30$^{o}$ in most of the cases. The proton energy 
spectra at the angles 10$^{o}$ and 50$^{o}$ are displayed in 
Fig.~\ref{alp1}~(right) for the reaction $^{20}$Ne (200 MeV) + $^{12}$C. 
It is evident that there is significant contribution from pre-equilibrium 
processes at forward angles, so far as proton emission is concerned. On 
the contrary, it is clear from Figs.~\ref{alp1}~(left) and \ref{alp2} that 
pre-equilibrium emission does not have any significant effect in the 
$\alpha$-particle emission spectra. The in-plane coincidence spectra of 
ER's (measured at $\theta_{lab}$ = 10$^{o}$) with LCP's (measured at 
$\theta_{lab}$ = 40$^{o}$) have been shown in 
Fig.~\ref{alp2}~(right pannel). It is seen that all the ER spectra peak 
at the energies corresponding to those of completely fused residues 
moving with velocity $\simeq$ $v_{CN} \cos \theta_{lab}$ (indicated 
by arrow). It is thus clearly evident that the $\alpha$-particles are
emitted predominantly from the evaporative decay of the hot composite.

\subsection{Invariant cross-section}
\begin{figure} [h]

{\epsfig{file=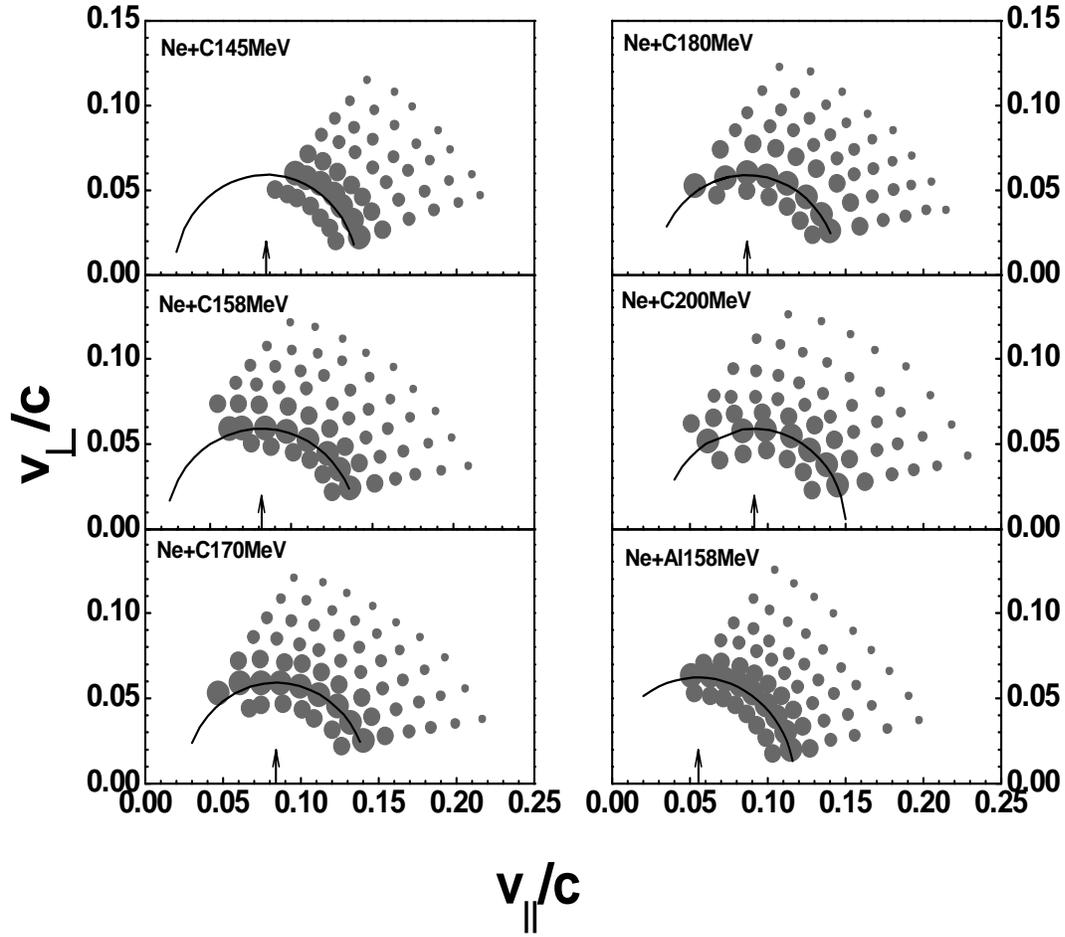,width=15cm,height=13.5cm.}}

\caption{ The invariant cross section of $\alpha$-particles plotted
in ($v_{\parallel}$,$v_{\perp}$) plane at different energies. The size of the 
point is in proportion to the invariant cross-section. The arrow corresponds 
to the compound nucleus velocity ($v_{CN}$) for
each energy. Solid curves are the most probable average velocities.}

\label{alp3}
\end{figure}

The velocity contour maps of the Galilean invariant differential
cross sections, {\it (d$^{2}$$\sigma$/d$\Omega$dE)p$^{-1}$c$^{-1}$},
as a function of the velocity of the emitted $\alpha$-particles
provide an overall picture of the reaction pattern. Fig.~\ref{alp3}
shows such velocity diagrams of invariant cross sections in the
($v_{\parallel}$,$v_{\perp}$) plane for the $\alpha$-particles
emitted at different incident energies. The arrows indicate the
compound nucleus velocity, $v_{CN}$. The circles correspond to the
most probable average velocities. It has been observed that for all
incident energies, the average velocities fall on a circle around
the compound nucleus velocity ($v_{CN}$). This implies that the
average velocities (as well as kinetic energies) of the $\alpha$'s
are independent of the center-of-mass (c.m.) emission angles. This
is an indication of complete energy relaxation and emission of
charged particles from a fully equilibrated source moving with
velocity $v_{CN}$.

\subsection{Statistical model analysis}
\begin{figure} [h]

{\epsfig{file=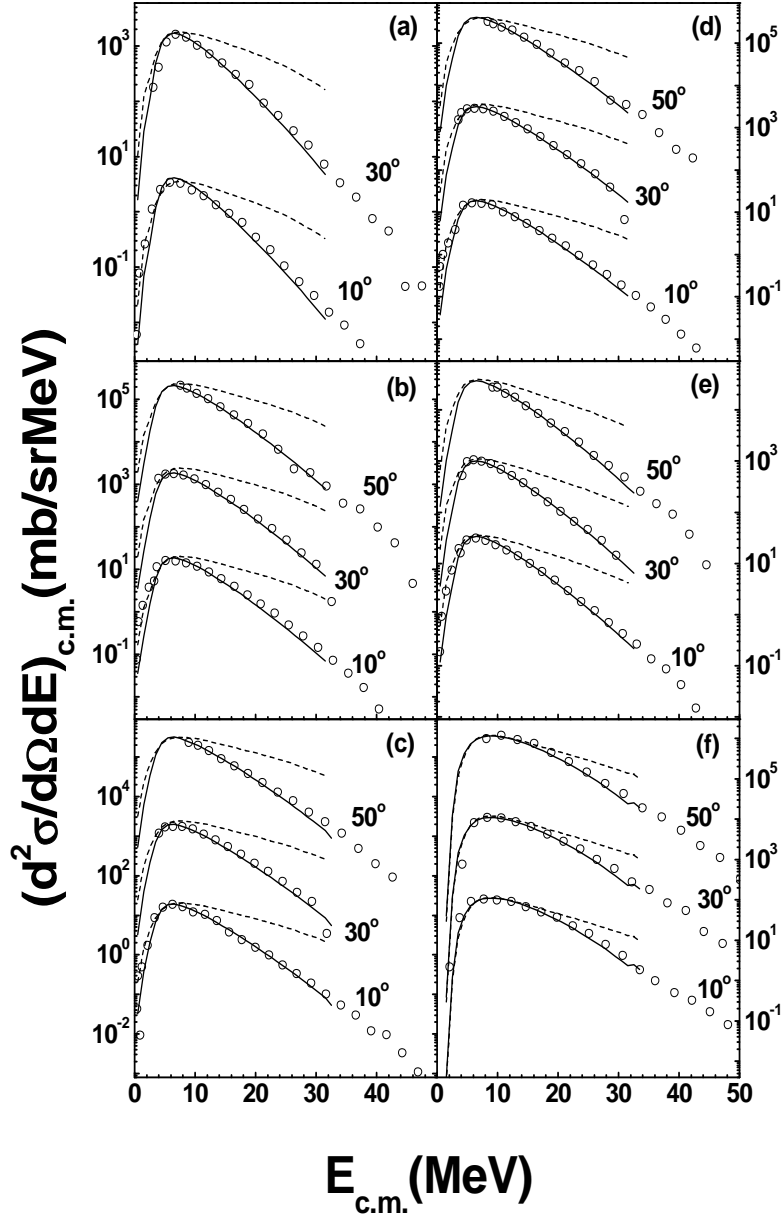,width=11cm,height=17.5cm.}}

\caption{ Inclusive energy spectra for $\alpha$-particles at
incident energies (a) 145 MeV, (b) 158 MeV, (c) 170 MeV, (d) 180 MeV and 
(e) 200 MeV for the system $^{20}$Ne + $^{12}$C and (f) 158 MeV for 
$^{20}$Ne + $^{27}$Al system at the laboratory angles 10$^{o}$ 
($\times$ 10$^{-2}$), 30$^{o}$ ($\times$ 10$^{0}$) and 50$^{o}$ 
($\times$ 10$^{2}$), respectively. Open
circles represent the experimental data. Dashed (solid) curves are the
predictions of the statistical model code CASCADE without (with)
deformation ({\it see text}).}

\label{alp4}
\end{figure}

The experimental energy spectra of $\alpha$-particles for a few
representative laboratory angles, {\it i.e.},
10$^{o}$ and 30$^{o}$ for 145 MeV and 10$^{o}$, 30$^{o}$ and 50$^{o}$ 
for other bombarding energies are displayed in Fig.~\ref{alp4}. The open 
circles represent the inclusive experimental data. At each emission 
angle the energy spectra exhibit approximately exponential slopes. 
The shapes and
high-energy slopes of the spectra are essentially independent of
center-of-mass emission angle. This is a signature of a
statistical de-excitation process arising from a thermalized source
and the exponential slope may be used to extract the source
temperature.

The analysis of the data ($^{20}$Ne + $^{12}$C) has been performed
using the standard statistical model code CASCADE \cite{ref10}. The 
dashed lines represent the predictions of the statistical model code 
CASCADE. The optical-model parameters used for calculating the
transmission coefficients were taken from Perry and Perry
\cite{ref11} for protons and from Huizenga and Igo \cite{ref12} for
$\alpha$-particles. The critical angular momentum for fusion, $j_{cr}$, 
was calculated using the Bass model \cite{ref13} for each bombarding 
energy. The radius parameter, $r_{o}$, was taken to be
1.29 fm following P\"uhlhofer {\it et al.} \cite{ref10}. The angular
momentum dependent deformation parameters ($\delta_{1}$ and
$\delta_{2}$ as described later) were set equal to zero \cite{ref6}. 
From Fig.~\ref{alp4}, it is clear that the theoretical calculations 
fail to predict the overall shape of the experimental spectra
over the whole energy range.

In the statistical model calculations, the lower energy part of the
LCP spectrum is controlled by the transmission coefficients. On the
other hand, the high energy part of the spectrum depends crucially
on the available phase space obtained from the level densities at
high spin. In hot rotating nuclei formed through heavy-ion
reactions, the nuclear level density at high angular momentum is taken 
to be spin dependent. For the calculation of transmission coefficients, 
we have used the optical-model parameters as mentioned earlier. A
deformed configuration of the compound nucleus has been assumed in
the calculation. This is realized by varying the radius parameter
$r_{o}$ as well as the deformation parameters $\delta_{1}$, $\delta_{2}$ 
of the spin dependent moment of inertia (as explained later in the text). 
The optimum value of $r_{o}$ is found to be $\sim$ 1.35 fm.
The increased radius parameter lowers the emission barrier and hence
enhances the low-energy yield by increasing the transmission
coefficients. The high-energy part of the spectrum is controlled by the 
level density $\rho(E^{*}, j)$. For a given angular momentum $j$ and
excitation energy $E^{*}$, the level density is defined as
\cite{ref6,ref10}:

\begin{equation}
\rho(E^{*}, j) = \frac{(2j + 1)}{12} a^{1/2}
\left(\frac{\hbar^{2}}{2{\mathcal{I}}_{eff}}\right)^{3/2}
\frac{1}{\left(E^{*} + T - \Delta - E_{j}\right)^{2}} exp\left[2
\left[a \left(E^{*} - \Delta - E_{j}\right)\right]^{1/2}\right],
\end{equation}

where, $a$ (taken as A/8, A being the mass number of the hot
rotating nucleus) is the level density parameter, T is the
thermodynamic temperature, $\Delta$ is the pairing correction, and,
$E_{j}$ is the rotational energy, which can be written in terms of
effective moment of inertia $\mathcal{I}$$_{eff}$ as,

\begin{equation}
E_{j} = \frac{\hbar^{2}}{2{\mathcal{I}}_{eff}} j(j+1).
\end{equation}

In hot rotating nuclei, formed through heavy-ion reactions, the
effective moment of inertia is taken to be spin dependent and is
written as

\begin{equation}
{\mathcal{I}}_{eff} = {\mathcal{I}}_{o} \times (1 + \delta_{1}j^{2} 
+ \delta_{2}j^{4}),
\end{equation}

where, the rigid-body moment of inertia, $\mathcal{I}$$_{o}$, 
is given by,

\begin{equation}
{\mathcal{I}}_{o} = \frac{2}{5} A^{5/3} r_{o}^{2}.
\end{equation}

Non-zero values of the parameters $\delta_{1}$, $\delta_{2}$
introduce the spin dependence in the effective moment of inertia.
The relation (2) defines a region in the energy-angular momentum
plane (E-j plane) of allowed levels, which 
is bounded by the yrast line, the locus of the lowest
energy states corresponding to each angular momenta. It has been
observed in the previous experiments \cite{ref1,viesti,ref6,bhat01} that
the yrast line has to be modified by varying the radius and
deformation parameters, $r_{o}$, $\delta_{1}$, $\delta_{2}$, to
explain the experimental data.

\begin{figure} [h]

{\epsfig{file=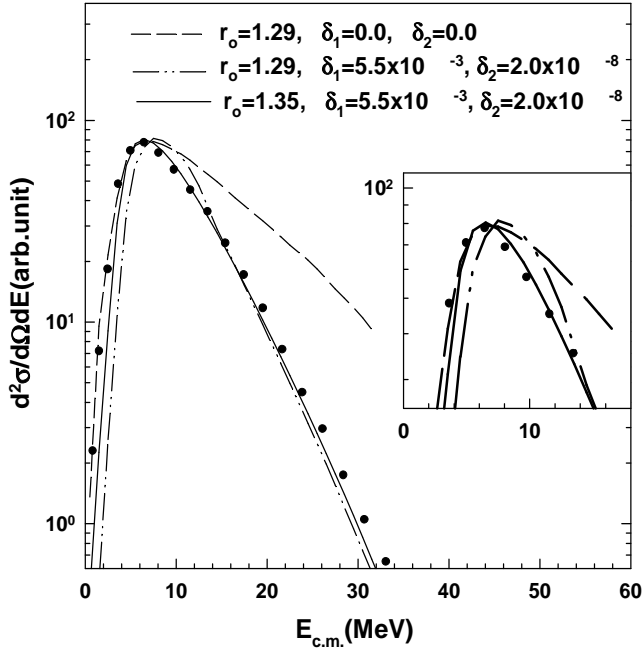,width=9cm,height=9cm.}}

\caption{ Experimental $\alpha$-particle energy spectrum (filled circles) at 
10$^{o}$ for $^{20}$Ne (200 MeV) + $^{12}$C reaction is compared with 
different CASCADE predictions. Variation near the peak 
of the distribution is shown in inset.}
\label{alp5}
\end{figure}

\begin{figure} [h]

{\epsfig{file=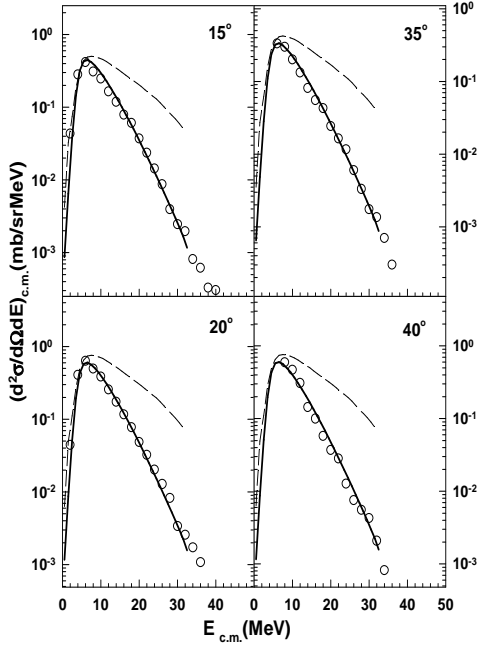,width=6.5cm,height=9cm.}}

\caption{ Exclusive $\alpha$-particle spectra at different
$\theta_{lab}$ measured in coincidence with ERs emitted at
$\theta_{lab}$ = 10$^{o}$ in $^{20}$Ne (158 MeV) + $^{12}$C reaction. 
Open circles are the experimental points,
dashed (solid) curves are  CASCADE predictions
without (with) deformation.}

\label{alp6}
\end{figure}

In the present work, the radius parameter, $r_{o}$, was varied from
1.29 fm to 1.35 fm to take care of the low energy side of the
experimental spectra. In addition, the variation of spin dependent
level density parameters was also needed for explaining the high
energy side of the spectra. This procedure is illustrated in 
Fig.~\ref{alp5} for $^{20}$Ne (200 MeV) + $^{12}$C reaction at an 
angle $\theta_{lab}$ = 10$^{o}$, where calculations with three 
different sets of parameters are displayed to highlight the roles 
played different parameters in explaining the data. 
It is clearly seen (see inset of Fig.~\ref{alp5}) that, the change of
$r_{o}$ from 1.29 fm to 1.35 fm causes a shift of the peak of the
distribution to low energy side and thus helps to fit the lower
energy part of the data in a better way. 
For different bombarding energies,
different sets of $\delta_{1}$, $\delta_{2}$ values were needed. 
The parameter $\delta_{1}$ plays the main role in explaining the
high energy part of the data; it is also found to be quite sensitive to
the variation of the excitation energy of the system. On the other hand,
non-zero value of the parameter $\delta_{2}$ is required to fit the
high energy tail ($\gtrsim$ 20 MeV) part of the data in particular, and
it is not found to be quite sensitive to the variation of excitation energy,
at least within the range of present measurements. Optimum values of
the parameters may, however, vary from system to system.
The optimized values of radius and deformation parameters are given in
Table~\ref{tbl2}. It is clearly seen that the experimental spectra
are well explained with these values (solid lines in Fig.
\ref{alp4}). The exclusive $\alpha$-particles spectra at
$\theta_{lab}$ = 15$^{o}$, 20$^{o}$, 35$^{o}$ and 40$^{o}$ are given
in Fig. \ref{alp6}. These $\alpha$-particles spectra were measured in
coincidence with ER's (10 $\leq$ Z $\leq$ 13). It is seen that the
exclusive spectra are also well explained with the same set of
parameters ($r_{o}$ = 1.35 fm, $\delta_{1}$ = 4.5 $\times$
10$^{-3}$ and $\delta_{2}$ = 2.0 $\times$ 10$^{-8}$) as those
obtained for the inclusive data at same energy.

From the above analyses, it is clear that large amount of spin
dependent deformation is needed to explain the emitted
$\alpha$-spectra from the $^{20}$Ne + $^{12}$C reaction. It will be
interesting to compare these deformations with those obtained for
other non $\alpha$-like light nuclei at similar excitation energy
and spin. Two representative systems [$^{47}$V$^{*}$ and
$^{31}$P$^{*}$, formed in $^{20}$Ne (158 MeV) + $^{27}$Al and
$^{19}$F (96 MeV) + $^{12}$C \cite{ref1}, respectively] were
considered for the purpose of comparison. The inclusive
$\alpha$-energy spectra for the decay of $^{47}$V$^{*}$ are shown in
Fig.~\ref{alp4} along with the CASCADE prediction. The optimized
values of the deformation parameters for both $^{47}$V$^{*}$ and
$^{31}$P$^{*}$ are given in Table~\ref{tbl2}. The deformation
parameters for both the systems are found to be very different from
those obtained for the decay of $^{32}$S$^{*}$.

A more quantitative comparison of effective deformations of hot
composites studied here can be made by examining the values of
effective radius parameters ($r_{eff}$) obtained from the analysis
of the data. The effective radius parameters (for non zero values of
$\delta_{1}$ and $\delta_{2}$ given in Table~\ref{tbl2}) were
estimated by averaging over the angular-momentum distribution (sharp
cutoff triangular distribution has been assumed, up to the critical
angular momentum, $j_{cr}$) in the following way:

\begin{equation}
r^{2}_{eff} = r^{2}_{o} \frac{\Sigma^{j_{cr}}_{o} (1 + \delta_{1}j^{2} 
+ \delta_{2}j^{4})(2j + 1)}{\Sigma^{j_{cr}}_{o} (2j + 1)}.
\end{equation}

The estimated values of the effective radius parameters are given in
Table~\ref{tbl2}. It is seen that $r_{eff}$ (and deformation)
increases with the increase in excitation energy. It is further
observed that values of $r_{eff}$ for $^{20}$Ne + $^{12}$C system at
all energies are significantly larger than those for the two
representative non-$\alpha$-like systems ($^{19}$F + $^{12}$C,
$^{20}$Ne + $^{27}$Al), even though in one case ($^{20}$Ne +
$^{27}$Al) the angular momentum involved is much larger. This
clearly indicates that the deformations of the hot composites formed
in $^{20}$Ne + $^{12}$C reactions are larger than those for other
light non-$\alpha$-like systems at similar excitation energies.

\section{Summary and Conclusion}
It has been observed that the $\alpha$-particles are
emitted predominantly from equilibrated compound nuclear sources in
all the reactions studied here. However, the experimentally measured
spectra deviate significantly from the standard statistical model
predictions. Significant amount of deformation is needed to explain
the measured spectra satisfactorily. A deformed configuration of the
compound nucleus is considered through the modification of the
radius parameter $r_{o}$ from 1.29 fm to 1.35 fm and by varying the
spin-dependent level density parameters $\delta_{1}$ and
$\delta_{2}$. It is evident that deformations needed to explain the
$\alpha$-spectra from $^{20}$Ne + $^{12}$C system are significantly
larger than those needed for $^{20}$Ne + $^{27}$Al and $^{19}$F +
$^{12}$C systems. Effective radius parameters extracted for the
$^{20}$Ne + $^{12}$C systems are also significantly larger than
those extracted for non-$\alpha$-like $^{19}$F + $^{12}$C and
$^{20}$Ne + $^{27}$Al systems -- indicating larger overall
deformations in the $^{20}$Ne + $^{12}$C systems. This may be
another signature of the formation of a long lived orbiting-like
dinuclear system at higher excitation energies ( $\sim$ 70 -- 100
MeV) in the $^{20}$Ne + $^{12}$C reaction, which was indicated
earlier \cite{ref9} from fragment emission studies.

\acknowledgements
The authors like to thank the cyclotron operating crew for smooth
running of the machine, and H. P. Sil for the fabrication of thin
Silicon detectors for the experiment. One of the authors (A. D.)
acknowledges with thanks the financial support provided by the
Council of Scientific and Industrial Research, Government of
India.

\begin{table}
\caption{ The optimized value of deformation parameters at each
incident energy.}
\begin{tabular}{ccccccc}
E$_{lab}$&E$^{*}$&$j_{cr}$&$r_{o}$&$\delta_{1}$&$\delta_{2}$&$r_{eff}$ \\
(MeV)&(MeV)&($\hbar$)&(fm)&&&(fm) \\ \tableline
145$^{a}$&73&24&1.35&3.7$\times$10$^{-3}$&2.0$\times$10$^{-8}$&1.95 \\
158$^{a}$&78&24&1.35&4.5$\times$10$^{-3}$&2.0$\times$10$^{-8}$&2.06 \\
170$^{a}$&82&24&1.35&5.0$\times$10$^{-3}$&2.0$\times$10$^{-8}$&2.12 \\
180$^{a}$&86&25&1.35&5.2$\times$10$^{-3}$&2.0$\times$10$^{-8}$&2.20 \\
200$^{a}$&94&25&1.35&5.5$\times$10$^{-3}$&2.0$\times$10$^{-8}$&2.24 \\
158$^{b}$&108&38&1.30&4.5$\times$10$^{-4}$&2.0$\times$10$^{-8}$&1.51 \\
96$^{c}$&60&21&1.29&2.8$\times$10$^{-3}$&2.5$\times$10$^{-7}$&1.65 \\
\end{tabular}
\label{tbl2}
\end{table}
\vspace{-0.4cm} $^{a}$ $\rightarrow$ $^{20}$Ne + $^{12}$C system,
$^{b}$ $\rightarrow$ $^{20}$Ne + $^{27}$Al system, $^{c}$
$\rightarrow$ $^{19}$F + $^{12}$C system \cite{ref1}

\newpage

\vspace{0.5cm}

\vspace{0.2cm}

\vspace{0.3cm}

\vspace{0.5cm}

\end{document}